\documentstyle[epsfig]{elsart}
\input psfig.sty
\def\frac#1#2{{#1\over#2}}
\def\be{\begin{equation}}
\def\ee{\end{equation}}

\begin{document}
\begin{frontmatter}
\title {
Effects of dust scattering 
in  expanding spherical nebulae}
\author [iia] {M. Srinivasa Rao \thanksref {msr}},
\author [iia] {Sujan Sengupta\thanksref{ssg}}

\thanks[msr]{E-mail: msrao@iiap.ernet.in}
\thanks[ssg]{E-mail: sujan@iiap.ernet.in}
\address [iia] {Indian Institute of Astrophysics, Bangalore 560034, India}

\begin{abstract}
The mean intensity of planetary nebulae with an expanding atmosphere is modeled
by considering dusty and dust-free atmospheres. The bulk matter density is
determined from the adopted velocity field through the equation of
continuity. The gas is assumed to consist
of hydrogen and helium and the gas-to-dust mass ratio is taken to be 
$3\times10^{-4}$. The Rayleigh phase function is
employed for atomic scattering while the full Mie theory of scattering
is incorporated for determining the dust scattering and absorption
cross-section as well as the phase function for the angular distribution
of photons after scattering.
It is shown that in a dust free atmosphere, the mean intensity increases
with the increase in the expansion velocity that makes the medium diluted.
The mean intensity profile changes significantly when dust scattering is
incorporated. The increase in forward scattering of photons by the dust
particles yields into an increase in the mean intensity as compared to that
without dust. The mean intensity increases as the particle size is increased.
Thus it is shown that both the expansion of the medium and the presence of
dust play important role in determining  the mean intensity of a planetary
nebulae.  
\end{abstract}

\end{frontmatter}
\section{Introduction}
There are several models of planetary nebulae which describe that the ionization
structure and other physical characteristics (Hjellming 1966, Rubin 1968,
Harrington 1968, 1969, Kirkpatrick 1970, 1972, Koppen 1979, 1980 ).
They calculated the ionization structure by solving simultaneously,
the equations of radiative transfer and ionization equilibrium for
several elements such as H, He, C, N, O, Ne, Mg, S, Ar etc. in different
stages of ionization together with the energy equilibrium equation. The ionization
equilibrium equation requires the use of the radiation in the form of mean intensity.
Normally mean intensity obtained by the `on the spot approximation' is utilized in
evaluating the formation of hydrogen Ly $\alpha$ line in dusty, expanding planetary 
nebulae. Infrared observations of planetary nebulae have established in the presence
of dust (like graphite, amorphous carbon, silicate, and iron) in these objects
and a comprehensive review work up to 1982 has been presented by Barlow (1983).
Pottasch et al (1984), Iyengar (1986), Zijlstra et al (1989), Ratag et al
(1990), Amnuel (1994). Pottasch et al. (1984) found infrared excesses
in high dust temperature nebulae. They concluded the possibility of
dust being mostly heated by the radiation of the central star on the long wave
side since the nebulae are young. Ratag et al. (1990) suggested that
the infrared excesses was due to the dust contents or heating by the
interstellar radiation field. Amnuel (1994) favored the latter explanation.
Peraiah and Wehrse (1978) found that the mean intensities of the radiation
field are far different from those obtained from the `on the spot
approximation'.  The expansion in the nebulae would certainly modify
the radiation field and this radiation field in tern changes the ionization 
structure in the nebulae. However,
a complete and rigorous analysis by incorporating relevant theory for
dust scattering in an expanding medium is not explored. In this paper we
investigate the changes that are produced on the radiation field by the 
radially expanding gases and dust.
For this purpose, we develop the solution of radiative transfer equation in the 
expanding nebulae. We assume both gas and dust with hydrogen and helium as the
components of the gas. In section 2 we present the relevant radiative transfer
equations that describes the radiation field under consideration. In section
3 we present the absorption cross section for atoms and dust. The adopted
velocity law, the density distribution and the dust parameters are presented
in section 4. The numerical method is described in section 5. The results are
discussed in section 6 followed by our conclusion in section 7.  

\section{Radiative Transfer Equations }

In a spherically symmetric, expanding, dusty planetary nebulae, the equations of
radiative transfer can be written (Peraiah and Grant, 1973; henceafter
PG73) as :
\begin{eqnarray}
&&\mu \displaystyle  \frac{\partial U(r, \mu)} {\partial r} +
\displaystyle \frac{1} {r} \displaystyle  \frac{\partial}
{\partial \mu} \biggl [(1-\mu^2) U(r, \mu)\biggr ]+
\sigma_g (r, \mu) U(r, \mu)
+\sigma_d(r, \mu) U(r, \mu)\nonumber \\ 
&&= \sigma_g (r, \mu) \biggl[\biggl(1-\omega_g(r)\biggr)B_g(r)
+\displaystyle  \frac{1} {2}\omega_g (r) \int^{+1} _{-1}
p(r, \mu, \mu^\prime)U(r, \mu^\prime)d\mu^\prime \biggr] \nonumber \\
&&+ \sigma_d (r, \mu) \biggl[\biggl(1-\omega_d(r)\biggr)B_d(r) 
+\displaystyle  \frac{1} {2}
\omega_d (r) \int^{+1} _{-1}p(r, \mu, \mu^\prime)U(r, \mu^\prime)d\mu^\prime \biggr]
\end{eqnarray}

fo the outward-going rays, and

\begin{eqnarray}
&&-\mu \displaystyle \frac{\partial U(r, -\mu)} {\partial r}
-\displaystyle  \frac{1} {r} \displaystyle \frac{\partial}
{\partial \mu} \biggl [(1-\mu^2) U(r, -\mu)\biggr ]
+\sigma_g (r, -\mu) U(r, -\mu) \nonumber \\
&&+\sigma_d (r, -\mu) U(r, -\mu) 
= \sigma_g (r, -\mu) \biggl[\biggl(1-\omega_g(r)\biggr)B_g(r)\nonumber \\
&&+\displaystyle  \frac{1} {2}\omega_g (r) \int^{+1} _{-1}
p(r, -\mu, \mu^\prime)U(r, \mu^\prime)d\mu^\prime\biggr]
+\sigma_d (r, -\mu) \nonumber \\
&&\biggl[\biggl(1-\omega_d(r)\biggl)B_d(r)
+\displaystyle  \frac{1} {2}\omega_d (r) \int^{+1} _{-1}        
p(r, -\mu, \mu^\prime)U(r, \mu^\prime)d\mu^\prime \biggr]
\end{eqnarray}
for  the inward-going rays.

\noindent The quantity $U(r, \mu)$ is given by,
\begin{eqnarray}
U(r, \mu)=4\pi r^2 I(r, \mu)
\end{eqnarray}
where $I(r, \mu)$ is the specific intensity of the ray making an angle
$\cos^{-1}\mu$
with the radius vector at radial points $r$. We have restricted
$\mu=\cos\theta$ to lie in the interval (0,1). Similarly,
\begin{eqnarray}
B_{g, d}(r)=4\pi r^2 b_{g, d}(r)
\end{eqnarray}
where the quantities $b_{g, d}(r)$ are Planck functions for gas and dust respectively.
$\omega_g$ and $\omega_d$ are the albedo's for single scattering for gas and dust.
The quantities $\sigma_g$ and $\sigma_d$ are the absorption cross-sections
for gas and dust.

\section{Absorption Cross-section for Gas and Dust}

 For the absorption cross-section of hydrogen and helium we have (Hummer and
Seaton 1964)
\begin{eqnarray}
\sigma_{H}(\nu) &=& 6.3 \times 10^{-18} \bigg ({\frac {\nu_0}{\nu}}\biggr) ^3 cm^2 \ \ \ \ \ \ 
 if \nu \geq \nu_0 \nonumber \\
&=& 0 \quad\quad\quad\quad\quad\quad\quad\quad\quad\quad  if \nu < \nu_0
\end{eqnarray}
for hydrogen and 
\begin{eqnarray}
&&\sigma_{He}(\nu) = 7.35 \times 10^{-18} 
exp \Biggr[-0.73 \Biggr({\frac{\nu_0}{\nu}}-1.808\Biggr)\Biggr] cm^2 \nonumber \\
&&\quad\quad\quad\quad\quad\quad\quad\quad\quad\quad  if \nu \geq 1.808 \ \nu_0 \nonumber \\
&&\quad\quad\quad = 0\quad\quad\quad\quad\quad\quad if \nu <  1.808 \ \nu_0 
\end{eqnarray}
for helium. Here $\nu_0$ is taken to be $3.289 \times 10^{15}$ Hz which is 
hydrogen Lyman limit.

The calculations are being done in the rest frame and therefore, we have to modify
the absorption cross section $\sigma_H$ and $\sigma_{He}$ in the rest frame by taking
into account of the Doppler effect. The equation (5) and the (6) can be
rewritten as (Landau and Lifshitz 1975)
\begin{eqnarray}
\sigma_{H}^+ (\nu)&=& 6.3 \times 10^{-18} 
\ \ \ \Biggr[\frac {\nu_0}{\nu}\Biggr(1+\frac{v}{c}\mu \Biggr)\Biggr]^3 cm^2,
\end{eqnarray}
\begin{eqnarray}
\sigma_{H}^- (\nu)&=& 6.3 \times 10^{-18} 
\ \ \ \Biggr[\frac {\nu_0}{\nu}\Biggr(1-\frac{v}{c}\mu \Biggr)\Biggr]^3 cm^2
\end{eqnarray}
if \quad  $\min \quad \Biggr[\nu \Biggr(1\pm \frac {v}{c}\mu\Biggr)\Biggr]\geq\nu_0$ 
 \quad   and 
\begin{eqnarray}
\sigma_{H}^\pm (\nu)&=& 0 \quad
\end{eqnarray}
  if $ \max \Biggr[\nu \Biggr(1\pm \frac {v}{c}\mu\Biggr)
\Biggr]<\nu_0 $
for hydrogen. For helium, 
\begin{eqnarray}
&&\sigma_{He}^+(\nu) = 7.35 \times 10^{-18} 
exp \Biggr[-0.73 \Biggr\{ 
\bigl(\nu+\frac {v} {c}\mu\bigr)
\frac {1} {\nu_0}-1.808\Biggr\}\Biggr] 
\end{eqnarray}
and
\begin{eqnarray}
&&\sigma_{He}^-(\nu) = 7.35 \times 10^{-18}
exp \Biggr[-0.73 \Biggr\{ 
\bigl(\nu-\frac {v}{c}\mu \bigr) 
\frac{1}{\nu_0}-1.808\Biggr\}\Biggr] 
\end{eqnarray}

if\quad  $\min \Biggr[\nu \Biggr(1\pm \frac {v}{c}\mu\Biggr)\Biggr]\geq 1.808 \ \nu_0$,
\begin{eqnarray}
\sigma_{He}^\pm(\nu)=0 \quad
\end{eqnarray}
 if $ \max \Biggr[\nu \Biggr(1\pm \frac {v}{c}\mu\Biggr)\Biggr]< 1.808 \ \nu_0$,
where $v$ is the radial velocity and $c$ is the velocity of light. We have restricted our 
calculations up to a maximum frequency equal to $\nu=4 \times \nu_0$. From equations
(7), (8), (10) and (11) it is clear that
\begin{eqnarray}
\sigma_g(r, \mu, \nu)=\sigma^+_g(r, \nu)=\sigma ^+_H(r, \nu)+\sigma^+_{He}(r, \nu) \nonumber \\
\sigma_g(r, -\mu, \nu)=\sigma^-_g(r, \nu)=\sigma ^-_H(r, \nu)+\sigma^-_{He}(r, \nu)
\end{eqnarray}
We have employed Rayleigh phase function that governs the angular distribution
of photons after scattering with the atoms and is given by (Chandrasekhar 1960) 
\begin{eqnarray}
P(r, \mu, \mu^\prime) =\frac{3}{8}[3-\mu^2+(2\mu^2-1)\mu'^2]
\end{eqnarray}

For dust absorption and scattering we have employed the Mie theory of
scattering (van de Hulst 1957). The extinction and scattering cross-section
for dust scattering are given by
\begin{eqnarray}
\sigma_d^{ext}=\frac{2\pi a^2}{x^2}Real\left[\Sigma_{n=1}^{\infty}(2n+1)
(a_n+b_n)\right]
\end{eqnarray}
and
\begin{eqnarray}
\sigma_d^{scat}=\frac{2\pi a^2}{x^2}\Sigma_{n=1}^{\infty}(2n+1)
(|a_n|^2+|b_n|^2)]
\end{eqnarray}
where $a$ is the radius of dust particles and $x=2\pi a/\lambda$. The Mie
coefficient $a_n$ and $b_n$ are in general complex function of the refractive
index of the dust and of $x$. We have calculated the dust extinction and
scattering cross section and hence the albedo $\omega_d$ numerically for
specific values of the dust size and refractive index. We have employed the
full Mie phase function for the angular distribution of photons after 
scattering with dust.

\section{Model Parameters}
We have adopted the usual power law for velocity profile which is given as
\begin{eqnarray}
V(r)=V_0+V_t\left(1-\frac{R_{in}}{r}\right)^{\gamma},
\end{eqnarray}
where $V_0$ is the initial velocity at the inner radius $R_{in}=9.2568 \times 10^{16}$ cm,
and $V_t$
is the terminal velocity at $R_{out}=2.221632 \times 10^{17}cm$  taken from 
Osterbrock (1974,
page 135)
. We fix the initial velocity $V_0$ at 0.1 kms$^{-1}$.
The terminal velocity $V_t$ is considered to be 10, 20, 30, 50, 100 and 300
kms$^{-1}$.

The distribution of total density comes from the equation of continuity and
is written as
\begin{eqnarray}
\rho(r)=\frac{\dot M}{4\pi r^2 V(r)}
\end{eqnarray}
where $\dot M$ is the rate of mass loss. Since the particle number density
in planetary nebulae ranges from 0.1 to 100 cm$^{-3}$, we have set 
$\dot M=10^{-6}$ M$_\odot$yr$^{-1}$. The number density of He is taken to
be 0.1 times that of H.

The dust-to-gas mass ratio for IC 418 and NGC 7662 are derived by Hoare (1990)
to be $6\times10^{-4}$ and $3\times10^{-4}$ respectively.
In our model we have taken dust-to-gas mass ratio as $3\times 10^{-4}$.
It is generally believed that dust in planetary nebula is carbon based one
such as graphite. However, observation at infrared reveals the presence of
silicates (Stasinska \& Szczerba 1999; Pottasch 1987). Thus,  we  consider
silicate type grains to demonstrate the effect of forward scattering.
Now both the real and the imaginary refractive index for silicate grains are
function of frequency and they depend on the grain composition. The real part
of the refractive index for magnesium silicates such as Forsterite and
Enstetite varies from 1.55 to 0.8 while the imaginary part varies from
0.85 to 0.3 (Scott \& Duley 1996) in the frequency interval we have 
considered in the present work. For the sake of
simplicity, in the present work we have  taken an average value of the
refractive indices for all the freequency points. The real part of the
refractive index is taken to be 1.2 while the imaginary part is set at 0.5 for
all frequency points. We have assumed same
kind and size of spherical dust particles with radius 0.1 and 0.5 micron. 
It is worth mentioning that the Mie phase function reduces to that of
Rayleigh when the ratio of the particle size to the wavelength is very
small. Mie scattering is asymmetric in the sense that the forward scattering
and the backward scattering are not the same while Rayleigh scattering is
symmetric. The asymmetry parameter $g=<\cos\theta>$ is given by
(Bohren \& Huffman 1983)
\begin{eqnarray}
\frac{1}{4}x^2Q_{sca}<\cos\theta>&=&\sum_n\frac{n(n+2)}{n+1}Re\{a_na_{n+1}^*
+b_nb_{n+1}^*\}+ \nonumber \\ && \sum_n\frac{2n+1}{n(n+1)}Re\{a_nb_n^*\}.
\end{eqnarray}
where $Q_{sca}$ is the scattering efficiency and $*$ denotes the complex
conjugate.
In fig~1, we present the asymmetry parameter for different frequency points
with grain radii 0.1 $\mu m$ and 0.5$\mu m$. For Rayleigh scattering,
$g=0$ because of symmetry. If the particle scatters more light toward the
forward direction then g is positive while g is negative if the scattering
is directed more toward the back direction.  Although, the dust sizes 
considered is very small, the wavelength range that we are concerned is
much shorter than the particle radius. As a consequence, g is positive in all
frequency points considered in the present work. The numerical value of g
very near to +1 implies that the particles scatter
light mostly in the forward direction. For a given frequnecy point the
asymmetry increases as the particle size increases. On the other hand, for
a given particle size, the asymmetry increases as the frequency increases.
However, the asymmetry parameter depends on the refractive index as well.
With the increase in the refractive index, the absorbity of the particle
increases.  Fig~1 shows that g is almost constant with the particle radius
0.5 $\mu m$. When the imaginary part of the refractive index is decreased
from 0.5 to as small as 0.02, g increases with the increase in frequency
and saturates at much higher value.
The asymmetry parameter for different size of graphite grains is presented
by Draine (1985). The Mie scattering and extinction cross-sections for
astronomical silicates have been discussed in details by Draine \& Lee (1984).
At large value of $x$, the extinction efficiency $Q_{ext}$ approaches to 2.0.
It is worth mentioning here that the dust opacity 
plays dominant role over the phase function in determining the emergent flux.
 \begin{figure*}
{\epsfxsize=14cm \epsfysize=14cm \epsfbox{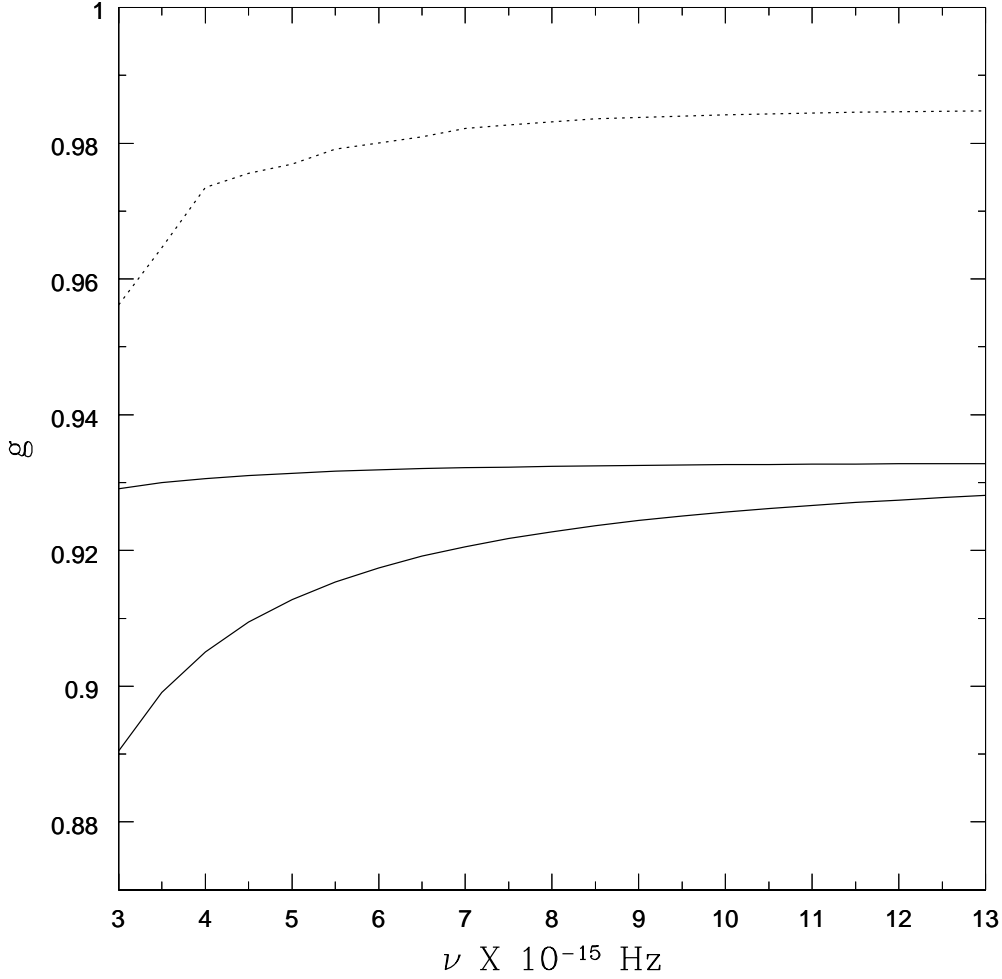}}
\caption{ The asymmetry parameter g with respect to the frequency.
Top solid line represents the value of g for the particle radius 0.5 $\mu m$
while bottom solid line represents that for particle radius 0.1 $\mu m$. The
dashed line represents the value of g for particle radius 0.5 $\mu m$ but with
imaginary refractive index 0.02}
\end{figure*}

\section{Numerical Method}

The solution of radiative transfer in spherical symmetry (1) and (2) 
is developed by using discrete space
theory of radiative transfer (PG73). In general the following
steps are followed for obtaining the solution. 

\noindent (i) We divide the medium into a 
number of "cells" whose thickness is less than or equal to the critical $(\tau_{crit})$.
The critical thickness is determined on the basis of physical characteristics of the medium. 
$\tau_{crit}$ ensures the stability and uniqueness of the solution.

\noindent (ii) Now the   integration  of the transfer equation is performed
on the "cell" which is two-dimensional radius - angle grid bounded by $[r_n, r_{n+1}] \times
[\mu_{j-\frac {1}{2}}, \mu_{j+\frac {1}{2}}]$ where $\mu_{j+\frac {1}{2}}$=
$\sum_{k=1}^{j} C_k, j=1,2 \ldots, J$, where $C_k$ are the weights of Gauss Legendre formula.

\noindent (iii) By  using the interaction principle described in PG73,
we obtain the reflection and transmission operators over the "cell".

\noindent (iv) Finally we combine all the cells by the star algorithm
described in PG73 and obtain the radiation field.  

Now equation (1) and (2) can be rewritten upon $\mu$ integration
\begin{eqnarray}
&&M \displaystyle \frac{\partial U^+(r)} {\partial r} + 
\frac{1} {r}\bigg [\Lambda^+U^+(r)+\Lambda^-U^-(r)\bigg] 
+\sigma_g^+ (r)U^+(r) \nonumber \\ 
&=& \sigma_g^+\bigg [1-\omega_g(r)\bigg]B_g^+(r) 
+\frac{1}{2} \omega_g \bigg [P^{+ + } (r)CU^+(r)
+ P^{+-}(r)CU^-(r)\bigg]\nonumber \\ 
&&+\frac {1} {2} \sigma_s \bigg [P^{+ + } (r)CU^+(r) 
+ P^{+-}(r)CU^-(r)\bigg] 
\end{eqnarray}
and
\begin{eqnarray}
&&-M \displaystyle \frac{\partial U^+(r)} {\partial r} - 
\frac{1} {r}\bigg [\Lambda^+U^+(r)+\Lambda^-U^+(r)\bigg] 
+\sigma_g^- (r)U^-(r) \nonumber \\
&=& \sigma_g^-(r)\bigg [1-\omega_g(r)\bigg]B_g^-(r) 
+\frac{1}{2} \omega_g (r)\bigg [P^{- + } (r) CU^+(r)
+ P^{--}(r)CU^-(r)\bigg] \nonumber \\
&&+\frac {1} {2} \sigma_s \bigg [P^{- + } (r)CU^+(r) + P^{--}(r)CU^-(r)\bigg] 
\end{eqnarray}

$\omega_g$ and $\omega_d$ are the albedo's for single scattering.
Here, $C$, $M$, $\sigma_g^-$ are the 
diagonal matrices of quadrature weights, angles and absorption cross-sections
corresponding to the set of $\mu$'s. $\Lambda ^+$ and $\Lambda^-$ are the curvature
matrices defined in PG73, and  

\begin{equation}
\bf U^\pm(r)= 
\left(\begin{array}{c}
U(r, \pm\mu_1) \\
U(r, \pm\mu_2) \\
U(r, \pm\mu_3) \\
\ldots \\
U(r, \pm\mu_J) \end{array}\right)
\end{equation}
and
\begin{eqnarray}
P^{++}(r)=P(r, +\mu, +\mu^\prime) 
\end{eqnarray}
where $0<\mu_j<\mu_J\leq 1$.
We discretize equations (19) and (20) by integrating these equations from $r_n$ to
$r_{n+1}$.
This gives us
\begin{eqnarray}
M\bigg[U^+_{n+1} - U^+_n\bigg] &+& \tau^+_{g, n+\frac{1}{2}} U^+_{n+\frac{1}{2}}
=\tau^+_{g, n+\frac{1}{2}} \bigg(1-\omega_g\bigg)B^+_{n+\frac{1}{2}}\nonumber \\
&&+\frac {1}{2}
\bigg(\tau_d I+\tau^+_{g, n+\frac{1}{2}} \omega_{g, n+\frac{1}{2}}\bigg)
\bigg(P^{++} C U^+_{n+\frac {1}{2}}+P^{+-} C U^-_{n+\frac {1}{2}}\bigg) \nonumber \\
&&-\rho_c \bigg(\Lambda^+ U^+_{n+\frac {1}{2}}+\Lambda^- U^-_{n+\frac {1}{2}}\bigg)
\end{eqnarray}
and
\begin{eqnarray}
M \bigg[U^-_{n} - U^-_{n+1}\bigg] &+& \tau^-_{g, n+\frac{1}{2}} U^-_{n+\frac{1}{2}}
=\tau^-_{g, n+\frac{1}{2}} \bigg(1-\omega_g\bigg)B^-_{n+\frac{1}{2}}\nonumber \\
&&+\frac {1} {2}
\bigg(\tau_d I+\tau^-_{g, n+\frac{1}{2}} \omega_{g, n+\frac{1}{2}}\bigg)
\bigg(P^{-+} C U^+_{n+\frac {1}{2}}+P^{--} C U^-_{n+\frac {1}{2}}\bigg) \nonumber \\
&&-\rho_c \bigg(\Lambda^- U^+_{n+\frac {1}{2}}+\Lambda^+ U^-_{n+\frac {1}{2}}\bigg)
\end{eqnarray}
where the quantity with subscript $n+\frac {1}{2}$ represents the average over the 'cell'
bounded by the radii $r_n$ and $r_{n+1}$ (here, the 'cell' is defined as a shell with radial
boundaries $r_n$ and $r_{n+1}$ and whose optical thickness is less than, or equal to 
the critical optical depth for which a stable and unique solution exists).
Here,
\begin{eqnarray}
&\Delta r_{n+\frac{1}{2}} = r_{n+1} -r_n \nonumber \\
&\tau^+_{g, n+\frac{1}{2}} = \sigma^+_{g, n+\frac{1}{2}} \Delta r_{n+\frac{1}{2}}
\end{eqnarray}
and $\tau_d$ is parametrized. The quantity $\rho_c$ is the curvature factor given by,
\begin{eqnarray}
\rho_c = \frac {\Delta r_{n+\frac{1}{2}}}{r_{n+\frac{1}{2}}}, 
\end{eqnarray}
where 
$
r_{n+\frac {1}{2}}=\frac {1} {2} (r_n+r_{n+1})
$
is suitable mean radius and 
$I$ is the identity matrix. The quantities $ U^\pm_{n+\frac{1}{2}}$ are approximated
by those at the boundaries $r_n$ and $r_{n+1}$ by (see PG73)
\begin{eqnarray}
U^\pm_{n+\frac{1}{2}}=\frac {1}{2}\bigg(U^\pm_n+U^\pm_{n+1}\bigg) 
\end{eqnarray}
Substitution of equation (27) into equation (23) and (24) would give us the system of equations in 
the form of interaction principle and the result is written as

\begin{eqnarray}
\left(\begin{array}{cc}
X11    &   X12 \\
X21    &   X22 \\ \end{array}\right)
\left(\begin{array}{c}
U^+_{n+1} \\
\\
U^-_{n \ \ }  \end{array}\right)
&&=\left(\begin{array}{cc}
Y11    &   Y12 \\
Y21    &   Y22 \\ \end{array}\right)
\left(\begin{array}{c}
U^+_{n+1} \\
\\
U^-_{n \ \ }  \end{array}\right)\nonumber \\
&&+(1-\omega_g)
\left(\begin{array}{cc}
\tau_g^+ & B^+\\ 
\tau_g^- & B^-\\ 
\end{array}\right) 
\end{eqnarray}
where
\begin{eqnarray}
&X11=
M+\frac {1}{2}\tau^+_g-\frac {1}{2}T^+P^{++}C+\frac {1}{2} \rho_c\Lambda^+ \nonumber \\
&X12=
-\frac {1}{2}T^+P^{++}C+\frac {1}{2}\rho_c\Lambda^- \nonumber \\
&X21=
 -\frac {1}{2}T^-P^{-+}C+\frac {1}{2}\rho_c\Lambda^-  \nonumber \\
&X22=
M-\frac {1}{2}\tau^+_g-\frac {1}{2}T^-P^{--}C-\frac {1}{2} \rho_c\Lambda^+ \nonumber \\
&Y11=
M-\frac {1}{2}\tau^+_g+\frac {1}{2}T^+P^{++}C-\frac {1}{2} \rho_c\Lambda^+ \nonumber \\ 
&Y12=
\frac {1}{2}T^+P^{+-}C-\frac {1}{2}\rho_c\Lambda^- \nonumber \\
&Y21=
\frac {1}{2}T^-P^{-+}C+\frac {1}{2}\rho_c\Lambda^-  \nonumber \\
&Y22=
M-\frac {1}{2}\tau^-_g+\frac {1}{2}T^-P^{--}C+\frac {1}{2} \rho_c\Lambda^+  \nonumber
\end{eqnarray}
\noindent where 
\begin{eqnarray}
T^\pm=\frac {1} {2}\bigg[\tau_d I+\tau_g^\pm \omega\bigg]
\end{eqnarray}
Comparing this with the interaction principle,
\begin{eqnarray}
\left(\begin{array}{c}
U^+_{n+1} \\
U^-_{n \ \ }  \\ \end{array}\right)& =&
\left(\begin{array}{cc}
t(n+1, n) & r(n, n+1) \\
r(n+1, n) & t(n, n+1) \\ \end{array}\right) \times 
\left(\begin{array}{c}
 U^+_{n \ \ } \\
U^-_{n+1} \\ \end{array}\right) 
+\left(\begin{array}{c}
\sum^+_{n+\frac{1}{2}} \\
\sum^-_{n+\frac{1}{2}} \\ \end{array}\right) 
\end{eqnarray}
where $t$'s and $r$'s are the transmission and reflection operators. The quantities
$\sum^{\pm}_{n+\frac {1}{2}}$ are the source vectors. By comparing equations (28) and (30)
one obtains the reflection and transmission operators. These are given in the appendix.

\section {Results and Discussion}

 \begin{figure*}
{\epsfxsize=14cm \epsfysize=14cm \epsfbox{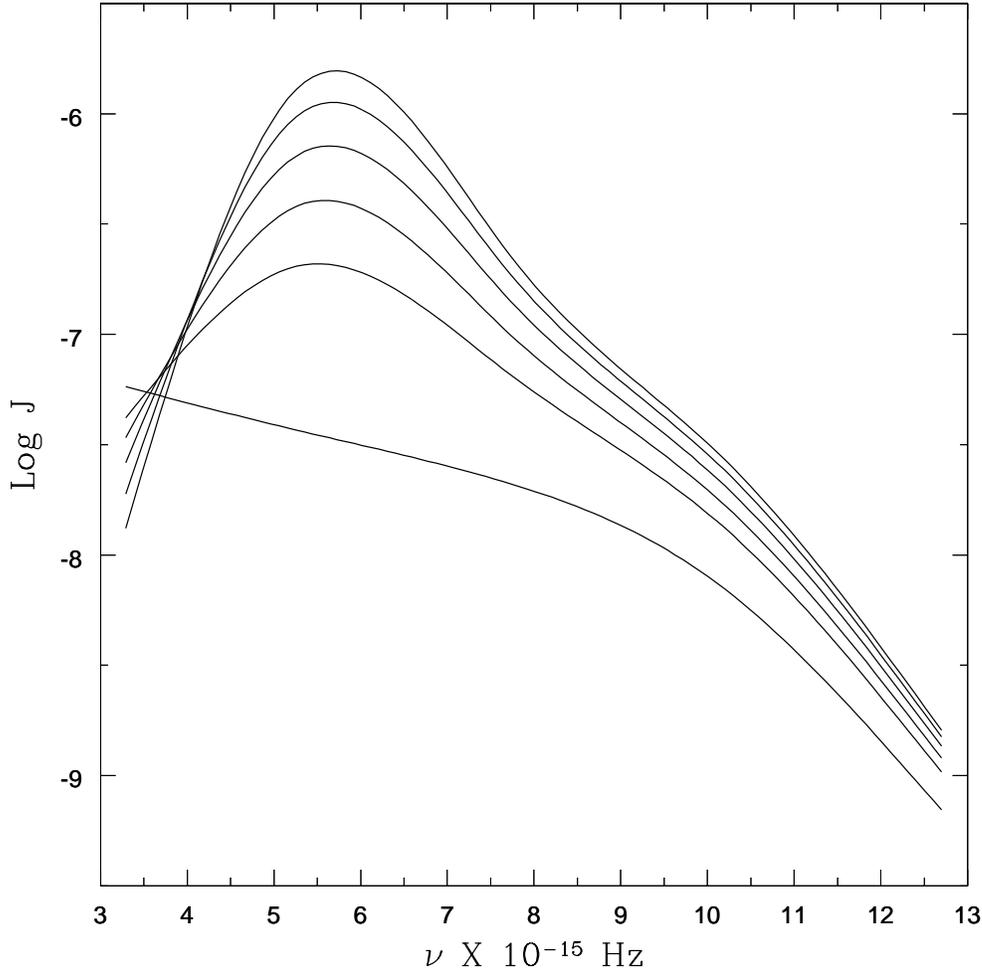}}
\caption{Mean intensity without dust particles for different frequency points.
From bottom to top the curves represent the models with terminal velocity
$V_t=$ 10, 20, 30, 50, 100 and 300 kms$^{-1}$ respectively. } 
\end{figure*}

 \begin{figure*}
{\epsfxsize=14cm \epsfysize=14cm \epsfbox{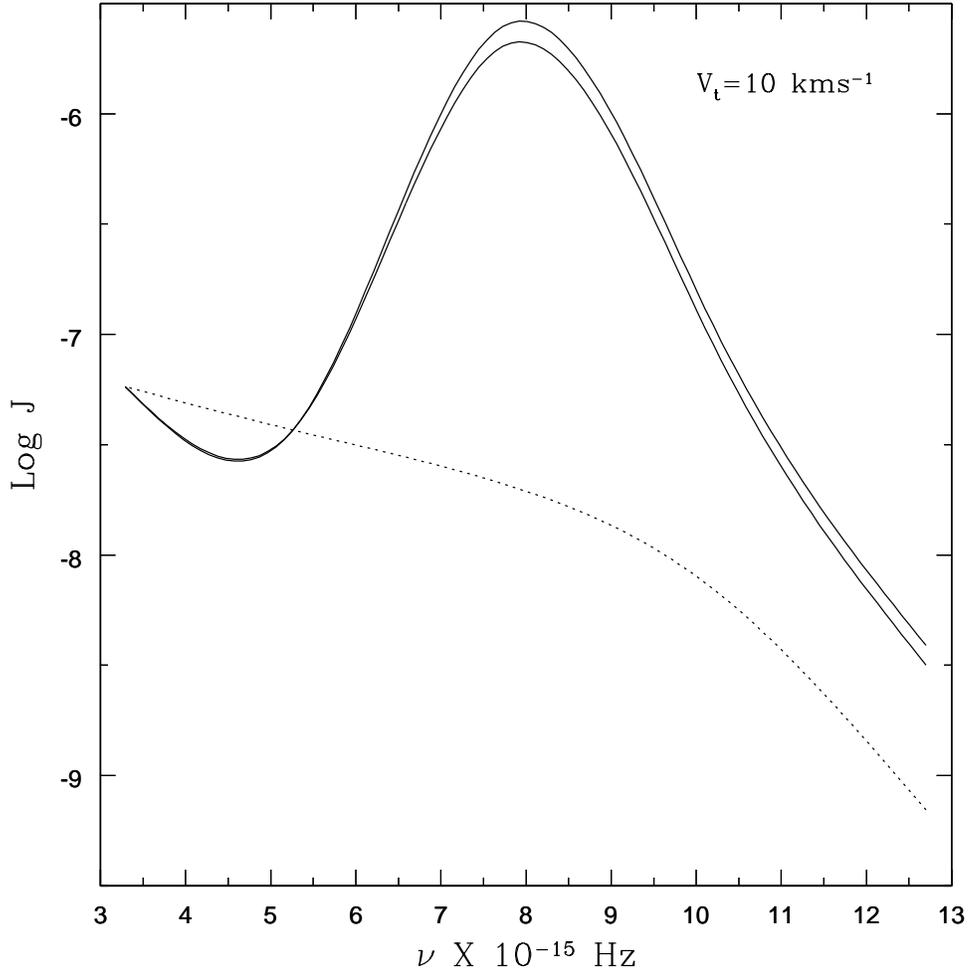}}
\caption{Mean intensity with dust particles of different sizes
and for $V_t$=10 kms$^{-1}$.
Top solid line represent the model with dust
particles of radius 0.5 $\mu m$ while the bottom solid line represent the
model with dust particles of radius 0.1 $\mu m$. The dashed line represents
the model without dust but with the same terminal velocity.}
\end{figure*}
 \begin{figure*}
{\epsfxsize=14cm \epsfysize=14cm \epsfbox{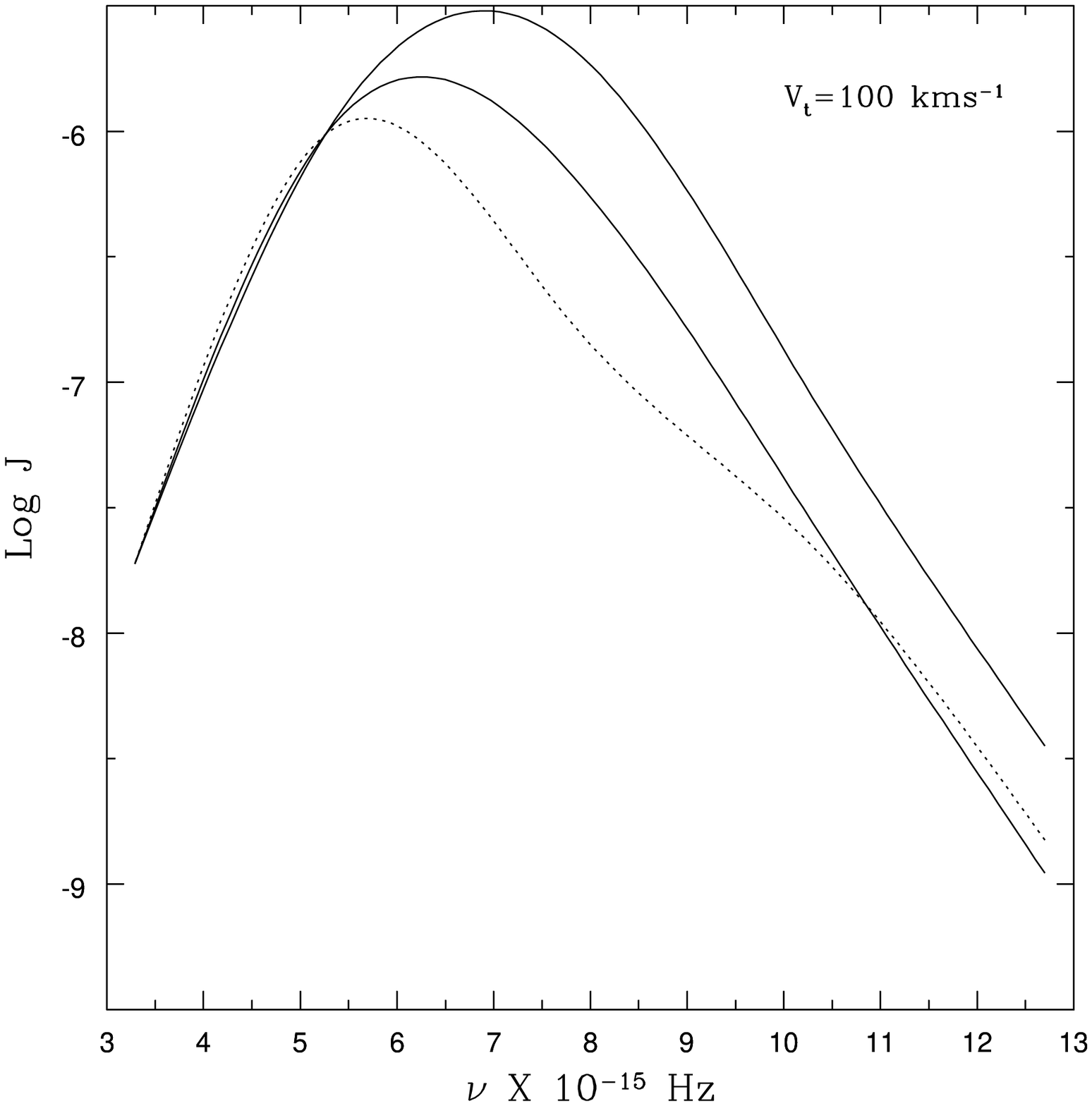}}
\caption{ Same as figure 3 but with $V_t$=100 kms$^{-1}$.}
\end{figure*}

 \begin{figure*}
{\epsfxsize=14cm \epsfysize=14cm \epsfbox{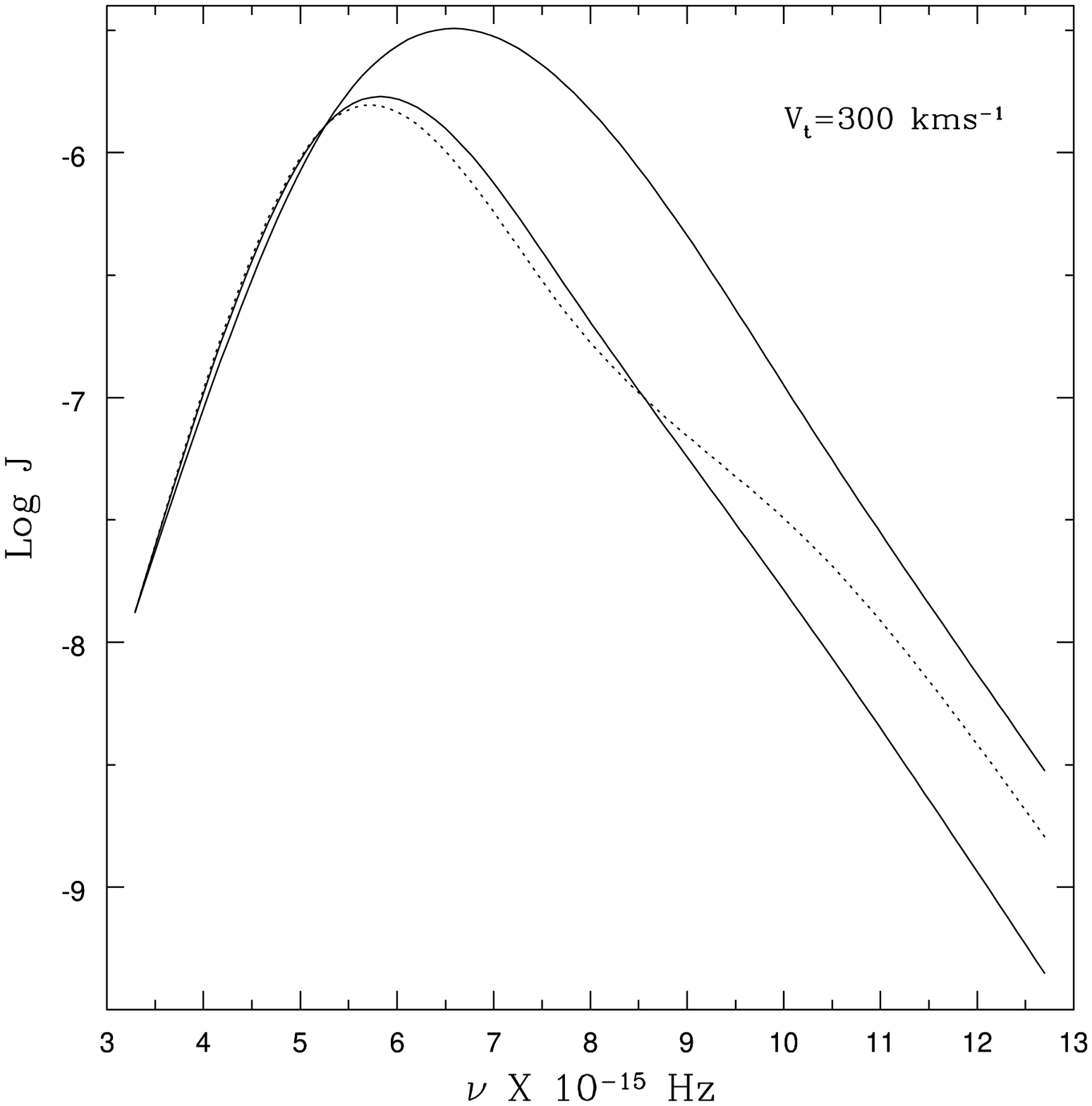}}
\caption{ Same as figure 3 but with $V_t$=300 kms$^{-1}$.}
\end{figure*}

As the purpose of this paper is to show the qualitative changes introduced by curvature and
expansion. We have chosen Harrington's (1969) model of NGC 7662 given in
Osterbrock (1974, page 135).
Few discrete points in the frequency grid are considered starting $3.289 \times 10^{15}$
Hz to about four times this frequency. We have divided this frequency interval into 20
points equally distributed along the grid. The mean intensity is calculated by
using the relation
\begin{eqnarray}
J_\nu=\frac {1}{2}\int ^{+1}_{-1} I(\mu) d\mu.
\end{eqnarray}

The results are presented in figure 2-5. Figure 2 shows the mean intensity
without the effect of dust scattering and absorption. The mean intensity
increases with the increase in the velocity. The change in the
mean intensity is maximum when the terminal velocity $V_t$ is increased
from 10 kms$^{-1}$ to 20 kms$^{-1}$. Expansion of the medium yields the
atmosphere diluted by  decreasing the density. Hence more photons can escape
from the bottom of the atmosphere. As a result, the mean intensity increases
with the increase in velocity. For a dust-free atmosphere, the opacity is
determined by the absorption cross section of Hydrogen and Helium. The
absorption cross section for Hydrogen varies with the cube of the frequency
while that for Helium changes exponentially. For small expansion velocity
the mean intensity falls slowly with the increase in frequency up to
$10^{16}$ Hz but then falls rapidly. This means the optical depth
with small velocity is mainly governed by the change in the Hydrogen opacity.
However, the Helium opacity starts dominating as the frequency increases from
$10^{16}$ Hz. When the velocity increases, the mean intensity increases
with the increase in frequency and peaks at about $5.5\times 10^{15}$ Hz
and then falls rapidly. The change in the qualitative feature of the mean
intensity with the increase in velocity is dictated by the change in the
absorption cross-section while the increase in the intensity is due to
the dilution of the medium.  
 
With the incorporation of dust grains, there are significant qualitative as
well as quantitative changes in the mean intensities as can be seen
from figure 3 to figure 5. This is due to the fact that in a dusty atmosphere
the optical depth is determined 
mainly by the dust. The absorption and scattering by dust are determined
by the size and the refractive index of the grains. In the present 
investigation we have considered dust grain with sizes 0.1 $\mu m$ and
0.5 $\mu m$. Given the high temperature of the medium, grains with larger
size would not survive. Since the wavelengths under consideration is
shorter than 0.1 $\mu m$, scattering by dust plays a significant role in
determining the mean intensity. It is well known that Rayleigh scattering
is symmetric so that the amount of forward and backward scattering is the same.
But, Mie scattering is asymmetric and it makes the forward scattering more
than the backward scattering as discussed in section 4. This asymmetry
increases as the ratio
between the wavelength and the particle size increases. In a dust free
clear atmosphere, a significant amount of radiation gets back scattered
contributing to the reduction in mean intensity. But, dust scattering makes
a large number of photons scattered in the forward direction yielding an
increase in the mean intensity as compared to that without dust. With the
adopted values for the wavelength and the particle size, dust scattering
dominates over dust absorption. As a result, increase in mean intensity is
obtained when dust is incorporated in the atmospheric models. The mean 
intensity increases with the increase in praticle size as the asymmetry
increases with the increase in particle size.
It is worth noting that the
mean intensity for two different sizes of dust particles is the same
for smaller frequency or longer wavelength and the changes in the mean
intensity increases as one moves from smaller to larger frequencies.

\section{Conclusion}
 
The mean intensity from a planetary nebulae with expanding atmosphere is
modeled by solving the radiative transfer equations for multiple scattering.
Both dusty and dust-free medium is considered. For the dusty medium, Mie
theory of scattering by spherical dust grains is employed. In a dust-free
atmosphere, the mean intensity increases with the increase in the expansion
velocity as the expansion of the medium removes matter causing a decrease
in optical depth. The entire intensity profile changes when dust scattering
is incorporated. Due to the increase in forward scattering of photon by dust
grains, the mean intensity increases significantly in a dusty medium as
compared to that in a dust-free atmosphere. Also, the mean intensity increases
with the increase in particle size.  Thus
it is shown that both the expansion and the presence of dust in the
atmosphere of planetary nebulae play important role in determining the
mean intensity of the object. The important implication of the present
work is that even the presence of  very small dust grain would affect
the spectrum of planetary nebulae at the ultra-violet and near optical
region. Therefore observation at shorter wavelengths would provide important
information on the properties of dust grains as well as the velocity field.

\ack
We are grateful to the anonymous referee for several useful suggestions and
comments that has not only improved the quality of the manuscript substantially
but also helped in removing some numerical errors.

\appendix 

\section [ ]{Appendix} 

The transmission and reflection operators as are given by,
\begin{eqnarray}
&t(n+1,n) = t^+\bigl[\Delta^+ S^{++} + r^{+-} r^{-+}\bigr]\nonumber \\ 
&t(n,n+1) = t^-\bigl[\Delta^- S^{--} + r^{-+} r^{+-}\bigr] \nonumber \\
&r(n+1,n) = 2t^- r^{-+} \Delta^+ M \nonumber \\
&r(n,n+1) = 2t^+ r^{+-} \Delta^- M 
\end{eqnarray}
and the source vectors are
\begin{eqnarray}
&\Sigma^+ = (1-\omega_g) \tau_g
t^+\bigl[ \Delta^+B^+ + r^{+-} \Delta^-B^-\bigr]\nonumber \\
&\Sigma^-= (1-\omega_g)\tau_g
t^-\bigl[ \Delta^-B^- + r^{-+} \Delta^+B^+\bigr]
\end{eqnarray}
where
\begin{eqnarray}
&\Delta^+ = \bigl[M + \frac{1} {2} \tau_g^+ (I-Q^{++})-\frac {1}{4}\tau_d P^{++}C\bigr]^{-1}\nonumber \\
&\Delta^- = \bigl[M + \frac{1} {2} \tau_g^- (I-Q^{--})-\frac {1} {4}\tau_dP^{--}C\bigr]^
{-1} 
\end{eqnarray}
\begin{eqnarray}
&S^{++} = M - \frac{1} {2} \tau_g^+ (I-Q^{++})+\frac {1} {4} \tau_d P^{++}C \nonumber \\
&S^{--} = M - \frac{1} {2} \tau_g^- (I-Q^{--})+\frac {1} {4} \tau_d P^{--}C 
\end{eqnarray}
\begin{eqnarray}
&S^{+-} = \frac{1} {2} \tau_d Q^{+-}-\frac {1}{4}P^{+-}C \nonumber \\
&S^{-+} = \frac{1} {2} \tau_d Q^{+-}-\frac {1}{4}P^{-+}C 
\end{eqnarray}

\begin{eqnarray}
&Q^{++} = \frac{1} {2} \omega P^{++} C-
\frac{\rho_c \Lambda^+} {\tau_g} \nonumber \\
&Q^{--} = \frac{1} {2} \omega P^{--} C+
\frac{\rho_c \Lambda^+} {\tau_g} \nonumber \\
&Q^{+-} = \frac{1} {2} \omega P^{+-}C-
\frac{\rho_c \Lambda^-} {\tau_g}
&Q^{-+} = \frac{1} {2} \omega P^{-+}C+
\frac{\rho_c \Lambda^-} {\tau_g} 
\end{eqnarray}
\end{document}